\documentclass[preprints,article,accept,moreauthors,pdftex]{mdpi} 
\usepackage{graphicx}
\usepackage{epstopdf}
\firstpage{1} 
\pubvolume{xx}
\issuenum{1}
\articlenumber{5}
\pubyear{2019}
\copyrightyear{2019}
\history{Received: 13 December 2019; Revised: 1 January 2020; Accepted: 7 January 2020}
\preto{\abstractkeywords}{\nolinenumbers}

\Title{Investigating Multiwavelength Lognormality with Simulations : Case of Mrk 421}

\Author{Nachiketa Chakraborty $^{1,2}$\orcidA{}*}

\AuthorNames{Nachiketa Chakraborty}

\address{%
$^{1}$ \quad Data Assimilation Research Centre, University of Reading, Whiteknights Rd., Reading RG6 6AH, UK ; n.chakraborty@reading.ac.uk\\
$^{2}$ \quad Max-Planck-Institut f\"ur Kernphysik, Saupfercheckweg 1, 69117 Heidelberg, Germany ; cnachi@mpi-hd.mpg.de}

\corres{Correspondence: cnachi@mpi-hd.mpg.de}

\abstract{Blazars are highly variable and display complex characteristics. A key characteristic is the flux probability distribution function or flux PDF whose shape depends upon the form of the underlying physical process driving variability. The BL Lacertae Mrk 421 is one of the brightest and most variable blazars across the electromagnetic spectrum. It has been reported to show hints of lognormality across the spectrum from radio to gamma-ray histograms of observed fluxes. This would imply that the underlying mechanisms may not conform to the "standard" additive, multi-zone picture, but could potentially have multiplicative processes. This is investigated by testing the observed lightcurves at different wavelengths with time-series simulations. We find that the simulations reveal a more complex scenario, than a single lognormal distribution explaining the multiwavelength lightcurves of Mrk 421.}


\begin{document}

\setcounter{section}{-1} 


\section{Introduction}

Blazars are the most extreme class of active galactic nuclei or AGNs \citep{2004ASPC..311...49U}. They are AGNs whose relativistic jets happen to be pointing towards the Earth such that the observed emission is relativistically Doppler boosted. Thus, they are very bright and are highly variable across the electromagnetic spectrum. Given that the jet is aligned with the line-of-sight of the observer, this emission is predominantly attributed to the jet. Therefore, any properties of the emission are either due to physical processes related to the jet or the ones that are propagated through from other components of the blazar such as the accretion disk through to the jet. The observed variability is over a very wide range of timescales ranging from years down to a few minutes. These variations are all modulated with noisy or stochastic variations that are physical and therefore of interest. While blazars exhibit very diverse features and behaviour in general in their variations, we can still determine certain trends that are more general. One key property in this respect is the observed flux distribution which is related to the probability distribution function (PDF) of the underlying stochastic processes. It encodes the form or class of physical process such as whether it is additive or multiplicative \citep[eg.,][]{1997MNRAS.292..679L, UMV:2005, 2018Galax...6..135R}. Thus by modeling PDF of individual sources, we can understand the physics of that particular source. And then combining results from different sources, we can potentially arrive at general trends of blazars. 

Mrk 421 is one of the closest blazars ($z = 0.031$) and has been studied in depth over many years across the electromagnetic spectrum \citep[eg.,][]{421logN}. It belongs to the sub-class of blazars called BL Lacertae objects and is one of the brightest in this category and also one of the most variable. It is a high frequency peaked blazar with a synchrotron emission peak at X-ray energies. There is also strong gamma-ray emission extending to TeV energies. It must also be noted that Mrk 421 has a significant contribution to the emission from the galaxy \citep{Gorham_2000}. Variability properties of Mrk 421 at different wavelengths have been derived from multiple observation campaigns listed before. \citet{421logN} observed this source at the full range of wavelengths ranging from radio with Ovens Valley Radio Observatory (OVRO) data to 
The High Altitude GAmma Ray (HAGAR) telescope system at gamma-ray energies. Amongst other observables, the flux probability distribution or the flux PDF was also estimated from all these lightcurves by histogramming the observed fluxes. As in a few other sources like PKS 2155-304 \citep{2155logN}, Mrk 501 \citep{2018Galax...6..135R}, they reported hints of lognormality along with a linear relation between the rms flux variations and the mean flux. From fits to the histogram of fluxes at every wavelength, they found a preference for fluxes to follow a lognormal distribution i.e., logarithm of the fluxes follow a normal distribution. This typically implies a multiplicative process. It must be noted that both the lognormality and linear relationship reported here is not equally significant and clear at all wavelengths. For instance, at radio wavelengths, the linear rms-flux relation is not necessarily the best fit. Additionally, the observed histogram was fit by a curve which is not the optimal way to estimate the PDF. In this paper, we test this scenario with time-series simulations. Once the shape of the PDF is determined, then one can model it with different classes of physical blazar emission models ranging from accretion disk driven fluctuations to local processes within the jet (eg synchrotron self-Compton or cascades). But we leave this modeling of physics for future work.

\section{Observations and Data}

The lightcurves used in this study were obtained from the published results in \citet{421logN} which were from a multiwavelength observational campaign on Mrk 421. This study was led by the HAGAR Telescope Array collaboration as a part of a regular monitoring campaign of Mrk 421. This campaign extended from 2009 to 2015 and provided a study of the long-term properties of the sources across the electromagnetic spectrum from radio to gamma-rays including TeV observations by HAGAR. We select from amongst the different lightcurves observed at different wavebands, Fermi-LAT at GeV energies, Swift-BAT in hard X-rays at keV energies and OVRO at GHz frequencies. These were selected as they have long and continuous coverage with the least amount of gaps. In principle, this could be extended to include the other lightcurves in 
\citet{421logN}. The radio data obtained from the OVRO are at 15 GHz from the publicly available lightcurves as a part of the Fermi blazar monitoring campaign \citep{2011ApJS..194...29R}. The radio lightcurves used for the variability analyses range from MJD $\sim$ 56266 to MJD $\sim$ 56792 a binning of 1 day. The Fermi lightcurves are in the range MJD $\sim$ 54840-56810 within the energy range $0.2 - 300$ GeV. The binning in this case is 10 days. The Swift-BAT lightcurves are observed between MJD $\sim$ 54840 to MJD $\sim$ 57180 within the energy range $15-150$ keV. The BAT lightcurves also have 10 day bins. All these selected lightcurves have the same number of data points, to enable an appropriate comparison of histograms. On the other hand, while the LAT and BAT lightcurves are essentially contemporaneous, OVRO lightcurves begin later. 
 \section{PDF Estimation Methodology}
 
While we can derive certain basic variability properties directly from lightcurves, including the histogram as an estimator of the underlying PDF, we need simulations for robust statistical results. Essentially as AGN lightcurves are stochastic, each lightcurve is a single realisation of this process. And hence, we need an ensemble of realisations to quantify the PDF. The PDF encodes both the spread of fluxes due to physics in addition to the uncertainties due to estimation and measurement. In order to generate an ensemble, we use spectral density estimation methods to generate the lightcurves. This is encoded in the popular method of Timmer and Koenig, \citep{TK:95}, hereafter referred to as TK95. This method generates gaussian or normally distributed time-series with power-law noise properties. We use parts of the \citet{2013MNRAS.433..907E} that generate TK95 lightcurves. We chose a power-law index, $\Gamma = 1$ or pink noise for our purposes. The nature of our conclusions is not highly sensitive to this choice as long as $\Gamma \leq 1.0$ as shown in \citep{2019MNRAS.489.2117M}. With a pink noise power spectral density (PSD) as input, we simulate lightcurves with the same time bins as the observed lightcurves but for a longer length. The longer length is the so-called red noise leakage factor \citep{2003MNRAS.345.1271V}and we set it to 5. By matching the time bins, we ensure that the exact cadence of the observations is replicated including gaps. This folds in the uncertainty and bias due to presence of gaps into our simulated ensemble and its spread.  As a result the simulations are realistic. Finally, as we want to explicitly test for lognormality, we exponentially transform the gaussian or normal TK95 simulations to one where the simulated fluxes follow a lognormal distribution in the same way as was done in \citet{2018Galax...6..135R}.
 
\section{Results : Normal vs Lognormal}
\begin{figure}
    \centering
    \includegraphics[width=0.45\textwidth]{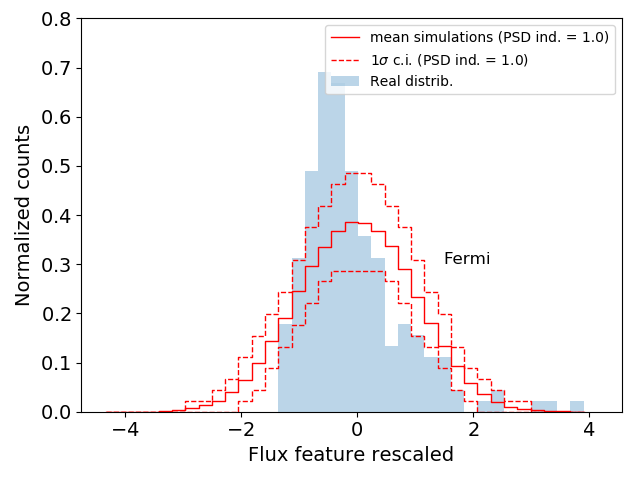}
    \includegraphics[width=0.45\textwidth]{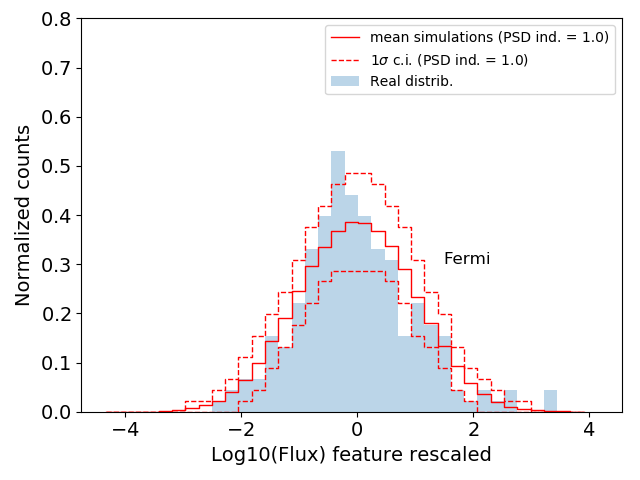}
    \caption{Figure shows the histogram of the observed Fermi-LAT lightcurves compared to those for simulations. Using simulations confidence intervals (dashed red) are derived at the 1$\sigma$ level for each bin. {\it Left}: Simulations of the fluxes that are normally distributed. {\it Right}: Simulations of the logarithm of fluxes that are normally distributed or lognormal simulations. The lognormal simulations fit the lightcurves significantly better with SW p-values of $.013$ relative to $1.78\times10^{-14}$ or $p \ll .001$ for the normal case.}
    \label{fig:fermipdf}
\end{figure}
 
Using the time-series simulations, we can statistically test the significance of the histogram bins. This is more robust than merely fitting a curve to the observed histogram. From the simulated lightcurves, we derive confidence intervals for each bin at the 1$\sigma$ confidence level. This is in effect a quantification of the the random, statistical error due to stochasticity of the fluxes. We generate simulations for both the normal and the lognormal cases. This is the done by exponentiating the normal simulations to derive the lognormal ones. This allows us to directly compare the two scenarios. 
\begin{figure}
    \centering
    \includegraphics[width=0.45\textwidth]{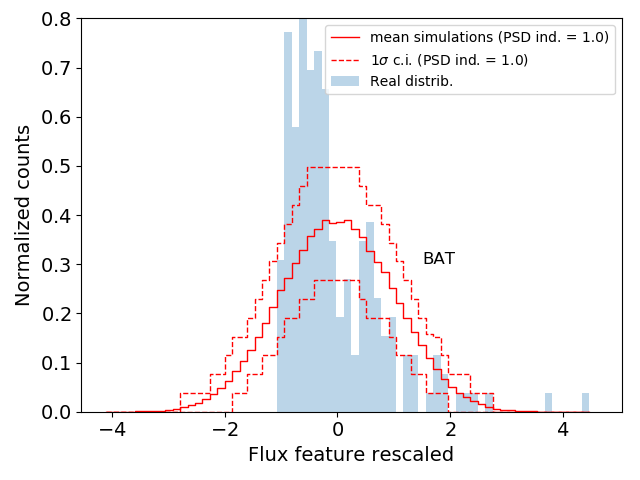}
    \includegraphics[width=0.45\textwidth]{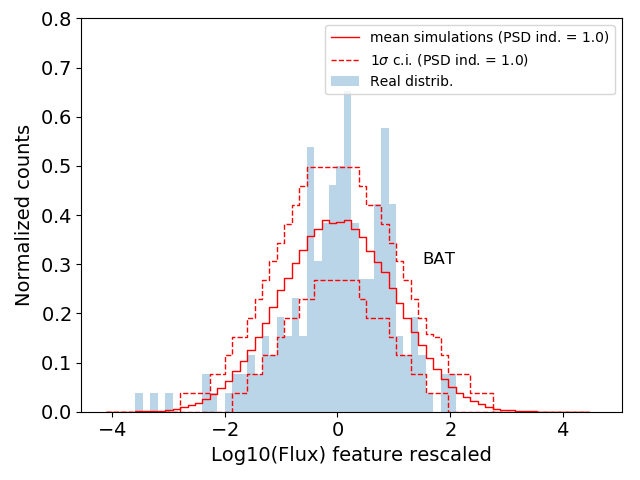}
    \caption{Similar to figure~\ref{fig:fermipdf} for BAT lightcurves. Both {\it left} normal simulations and {\it right} lognormal ones are not great fits with SW p-values of $8.29\times10^{-16}$ (i.e.$p \ll .001$) and $1.30\times10^{-6}$ (also, $p \ll .001$), even though the latter appears to be better. This mismatch is potentially due to the multi-modal, bursty structure.}
    \label{fig:batpdf}
\end{figure}

For Fermi-LAT lightcurves, we find that the lognormal simulations fit the data much better. There is a stronger statistical preference for the lognormality by the data, relative to a normal distribution. This is quantified in terms of the Shapiro-Wilks test \citep{swtest1965}. The Shapiro-Wilks test is most sensitive at rejecting the null hypothesis of a normal distribution. We apply this to both the fluxes and the logarithm of the fluxes. And we find that for the Fermi-LAT lightcurves, the flux values strongly reject the normal hypothesis with a p-value of $1.79\times10^{-14}$ or $p \ll .001$, whereas the Log(flux) values do not with a p-value of 0.013 as shown in the figure~\ref{fig:fermipdf}. This supports the lognormal scenario at a $99\%$ confidence level. Therefore, a single lognormal distribution does seem to describe the Fermi-LAT data adequately well. This suggests that the gamma-rays can be generated by a single process. 

\begin{figure}
    \centering
    \includegraphics[width=0.45\textwidth]{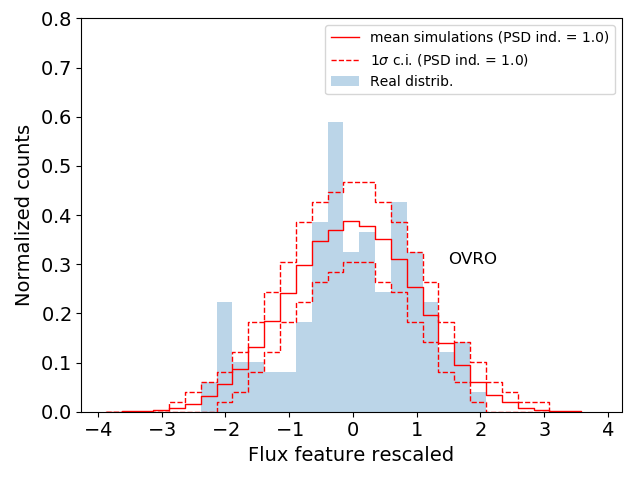}
    \includegraphics[width=0.45\textwidth]{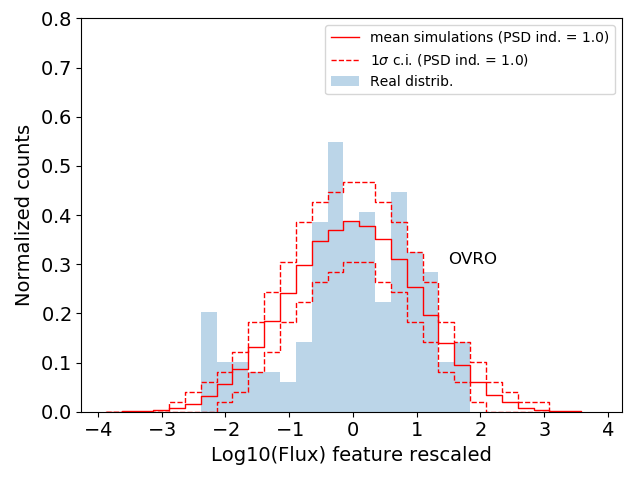}
    \caption{Similar to figure~\ref{fig:fermipdf} for OVRO lightcurves. Between {\it left} normal simulations and {\it right} lognormal ones, the normal distribution is a somewhat better fit with SW p-values of $2.31\times10^{-4}$ relative to $4.80\times10^{-6}$. Once again the multi-modal, bursty structure makes it complex to explain with a single model for PDF.}
    \label{fig:ovropdf}
\end{figure}

In case of the BAT lightcurves, the situation is less clear. It appears that the lognormal distribution "fits" the data much better. However, the SW p-values are $8.29\times10^{-16}$ and $1.30\times10^{-6}$ respectively which are both much less than .001. So naively, while one may conclude that the normal hypothesis is once again rejected over the lognormal, neither model is a great fit to the data. The absolute p-values suggest that the data are not sensitive to a clear discrimination. This is likely because the BAT lightcurves are bursty and even appear to have multiple modes. This is possibly why a simple, single lognormal model while capturing the tail of the flux distribution is still not a good enough model by itself. For OVRO lightcurves, this trend is reversed. The data seem to prefer a normal distribution to a lognormal distribution with a Shapiro-Wilks p-value of $2.31\times10^{-4}$ for the normal distribution and $4.79\times10^{-6}$ for lognormal distribution. It must be emphasised that in this case, the preference is not as strong the preference for lognormality in the Fermi-LAT case based on the numerical p-values and indeed on comparison of figures~\ref{fig:ovropdf} and ~\ref{fig:fermipdf}. Thus, statistically examining the observed histograms with simulated lightcurves mimicking the radio (OVRO), X-ray (Swift-BAT) and gamma-ray (Fermi-LAT) observations, we find that Mrk 421 is not an unambiguous case of MWL lognormality but shows more complex behaviour. 

\section{Discussion and Conclusions}

 The shape of the flux distribution is an important observable for AGNs and variable sources in general. It encodes the form of the physical processes. Lognormality for instance, is a natural outcome of a multiplicative process as stated earlier and in earlier research \citep[eg.,][]{1997MNRAS.292..679L, UMV:2005}. In particular, if there is lognormality observed across a wide bandwidth of electromagnetic spectrum (from radio to gamma-rays), then this suggests that there could be a cascade-type process at play. A highly energetic process producing gamma-ray variability, can successively dissipate energy to lower and lower energies which would be multiplicative due to its cascading nature. However, here we find that the different wavebands show different properties for the PDF. The Fermi-LAT lightcurves show lognormality, whereas the OVRO lightcurves have a bit of a preference for normal distribution. And the BAT lightcurves show more complex, possibly multi-modal behaviour with multiple bursts. This suggests multiple components, that cannot be explained by a simple model or a single distribution. Therefore, it is quite likely that the different wavebands are not all produced by a single cascade-like process, but in fact, multiple processes are at play in Mrk 421. This does not exclude a multiplicative cascade-like component, but this cannot be the dominant and certainly not the only source of variability in Mrk 421. Thus, we see that the time-series simulations can quantitatively discriminate between a single cascade-like process which produces all the multiwavelength emission and different processes driving variability at different wavelengths. We find that while directly from the data, deviations from normality including tails and bursts tend to favour lognormal distributions, we need to simulate to robustly determine the statistical significance of these results. 


\acknowledgments{NC kindly acknowledges the continued support from MPIK, Heidelberg and DARC, Reading for their resources to conduct this research. NC also thanks Atreyee Sinha for providing multiwavelength lightcurves that were published in a paper lead by her and the very useful discussions and parts of code from Carlo Romoli used to extract p-values from simulations performed by NC. Finally, NC thanks the insightful discussions with Daniela Dorner and Frank Rieger on the topic in general as well as the ano.}





\reftitle{References}
\externalbibliography{yes}
\bibliographystyle{mdpi}
\bibliography{pdfbib}



\end{document}